\documentclass{aa}

\usepackage{amssymb}
\usepackage{graphicx}

\begin{document}

\title{On the spectrum of high-energy cosmic rays produced by supernova 
remnants
in the presence of strong cosmic-ray streaming instability and wave 
dissipation}
\author{V.S. Ptuskin, \inst{1} \and V.N. Zirakashvili \inst{1,2}}
\date{Received ; accepted }

\titlerunning{On the spectrum of high-energy cosmic rays}

\offprints{V.S. Ptuskin}

\institute{Institute for Terrestrial Magnetism, Ionosphere and Radiowave
Propagation, 142190, Troitsk, Moscow Region, Russia\\
             \email{vptuskin@izmiran.rssi.ru}
         \and
Max-Planck-Institut f\"{u}r\ Kernphysik, D-69029,
Heidelberg, Postfach 103980, Germany\\
              \email{zirak@mpimail.mpi-hd.mpg.de}
             }

\abstract{
The cosmic-ray streaming instability creates strong magnetohydrodynamic
turbulence in the precursor of a SN shock. The level of turbulence 
determines
the maximum energy of cosmic-ray particles accelerated by the diffusive 
shock
acceleration mechanism. The present consideration continues our work Ptuskin
\& Zirakashvili (2003). It is assumed that the Kolmogorov type nonlinear 
wave
interactions together with the ion-neutral collisions restrict the amplitude
of random magnetic field. As a result, the maximum energy of accelerated
particles strongly depends on the age of a SNR. The average spectrum of
cosmic rays injected in the interstellar medium over the
course of adiabatic SNR evolution (the Sedov stage) is approximately
$Q(p)p^{2}\propto p^{-2}$ at energies larger than $10-30$ Gev/nucleon and 
with
the maximum particle energy that is close to
the knee position in cosmic ray spectrum observed at $\sim4\times10^{15}$ 
eV. At
earlier stage of SNR evolution - the ejecta dominated stage described by the
Chevalier-Nadyozhin solution, the particles are
accelerated to higher energies and have rather steep power-law distribution 
on
momentum. These results suggest that the knee may mark the transition from 
the ejecta-dominated
to the adiabatic evolution of SNR shocks which accelerate cosmic rays.

}

%________________________________________________________________
\maketitle

\section{Introduction}

The diffusive shock acceleration is considered as the main mechanism
of acceleration of galactic cosmic rays. The dependent on energy diffusion
coefficient $D(E)$ determines the maximum energy that particles can gain in
the process of acceleration by the shock moving through the turbulent
interstellar medium. The condition of efficient acceleration is $D(E)\leq
\varkappa u_{\mathrm{sh}}R_{\mathrm{sh}}$, where $R_{\mathrm{sh}}$ is the
radius and $u_{\mathrm{sh}}$ is the velocity of spherical shock, the 
constant
$\varkappa\sim0.1$, see Drury et al. (\cite{drury01}), Malkov \& Drury
(\cite{malkov}) for
review. The Bohm value of the diffusion coefficient $\ D_{B}=vr_{\mathrm{g}
}/3$ ($v$ is the particle velocity, and $r_{\mathrm{g}}$ is the particle
Larmor radius) that is a lower bound of the diffusion along the average
magnetic field, gives the maximum particle energy $E_{\max}\sim2\times
10^{14}Z\left(  \mathcal{E}_{51}/n_{0}\right)  $ eV at the time of 
transition
from the ejecta dominated stage to the stage of adiabatic evolution of SNR
(the particle charge is $Ze$). Here the SN burst with the kinetic energy of
ejecta $\mathcal{E=E}_{51}10^{51}$ erg in the gas with density $n_{0}$
cm$^{{-3}}$ and the interstellar magnetic field $B_{0}=5$ $\mu$G is
considered. The found value of $E_{\max}$ is close but somewhat less than 
the
energy of the ''knee'', the break in the total cosmic ray spectrum observed 
at
$\sim4\times10^{15}$ eV.

Analyzing the early stage of SNR evolution when the shock velocity is high,
$u_{\mathrm{sh}}\sim10^{4}$ km s$^{-1}$, Bell\ \& Lucek (\cite{bell01})
found that the
cosmic-ray streaming instability in the shock precursor can be so strong
that the amplified field
$\delta B\geq100$ $\mu$G far exceeds the interstellar value $B_{0}$. The
maximum particle energy increases accordingly. The cosmic-ray streaming
instability is less efficient as the shock velocity decreases with time and
the nonlinear wave interactions reduce the level of turbulence at the late
Sedov stage (V\"{o}lk et al. \cite{voelk}, Fedorenko \cite{fedorenko}).
This leads to the fast
diffusion and to the corresponding decrease of $E_{\max}$. The effect is
aggravated by the possible wave damping on the ion-neutral collisions (Bell
 \cite{bell78}, Drury et al. \cite{drury96}). The acceleration of cosmic 
rays and their
streaming instability in a wide range of shock velocities was considered in
our paper Ptuskin \& Zirakashvili (\cite{ptuskin03}) (Paper I). The 
analytical expressions
for cosmic ray diffusion coefficient and for the instability growth rate 
were
generalized to the case of arbitrary strong random magnetic field, $\delta
B\gtrless B_{0}$. The rate of nonlinear wave interactions was assumed to
correspond to the Kolmogorov nonlinearity of magnetohydrodynamic waves. The
collisional dissipation was also taken into account. The maximum energy of
accelerated particles was determined as a function of shock velocity and 
thus
as a function of SNR age. The maximum energy of the particle with charge 
$Ze$
can be as high as $10^{17}Z$ eV in some very young SNRs and falls down to
about $10^{10}Z$ eV at the end of adiabatic (Sedov) stage. The widely 
accepted
estimate of cosmic ray diffusion coefficient at the strong shock that
corresponds to the Bohm diffusion value calculated for the interstellar
magnetic field strength turns out to be not correct. This result may explain
why the SNRs with the age more than a few thousand years are not prominent
sources of very high energy gamma-rays (Buckley et al \cite{buckley},
Aharonian et al \cite{aharonian}). The presence of
strongly amplified random magnetic field in young SNRs is evidently 
supported
by the interpretation of data on synchrotron X-ray emission from young SNRs,
see e.g. Vink (\cite{vink}) for review.

The main objective of the present work is the calculation of the average
spectrum of cosmic rays ejected in the interstellar medium by a SNR in a
course of its evolution. Some necessary results of Paper I are presented in
the next Section 2, the evolution of SNR shocks is discussed in Section 3, 
the
average cosmic-ray source spectrum is calculated in Section 4
followed by the discussion in Section 5, the conclusion is given in Section 
6.
Appendix A describes the thin shell approximation used in our calculations.

\section{Maximum Energy of Accelerated Particles}

In the test particle approximation, the distribution of accelerated 
particles
in momentum for high Mach number shocks has the canonical form $f(p)\sim
p^{-4}$ (Krymsky \cite{krymsky}, Bell \cite{bell78}).
In the case of efficient acceleration, the
action of cosmic ray pressure on the shock structure causes nonlinear
modification of the shock that changes the shape of particle spectrum making
it flatter at ultra relativistic energies (Eichler \cite{eichler};
Berezhko et al. \cite{berezhko96},
Malkov \& Drury \cite{malkov}). Because of this effect, we assume that the
distribution of ultrarelativistic particles at the shock is of the form
$f_{0}(p)\sim p^{-4+a}$ where $0<a<0.5$, and the value $a=0.3$ is used in 
the
numerical estimates below. The normalization of function $f(p)$ is such that
the integral $N=4\pi\int dpp^{2}f(p)$ gives the number density of cosmic 
rays.
The differential cosmic ray intensity is $I(E)=f(p)p^{2}$. We assume that 
the
cosmic ray pressure at the shock is some fraction $\xi_{\mathrm{cr}}<1$ of 
the
upstream momentum flux entering the shock front, so that $P_{\mathrm{cr}}
=\xi_{\mathrm{cr}}\rho u_{\mathrm{sh}}^{2}$ and the equation for the
distribution function of relativistic accelerated particles at the shock is
\begin{equation}
f_{0}(p,t)=\frac{3\xi_{\mathrm{cr}}\rho 
u_{\mathrm{sh}}^{2}H(p_{\mathrm{\max}
}(t)-p)}{4\pi c(mc)^{a}\varphi(p_{\mathrm{\max}})p^{4-a}}, \label{1}
\end{equation}
where $p_{\mathrm{\max}}$ is the maximum momentum of accelerated particles,
$H(p)$ is the step function, and $\varphi(p_{\mathrm{\max}})=\int
_{0}^{p_{\mathrm{\max}}/mc}\frac{dyy^{a}}{\sqrt{1+y^{2}}}$. The 
approximation
of the last integral at $p_{\max}\gg mc$ is $\varphi(p)\approx a^{-1}
(p/mc)^{a}-a^{-1}(1+a)^{-1}$. The value of $\xi_{\mathrm{cr}}\approx0.5$ and
the total compression ratio at the shock close to $7$ were found in the
numerical simulations of strongly modified SN shocks by Berezhko et al.
(\cite{berezhko96}). Here and below we mainly consider protons as the most 
abundant cosmic
ray component. For ions with charge $Z$, the equations should be written as
functions of $p/Z$ instead of $p$. In particular, the nuclei with charge $Z$
reach the maximum momentum a factor of $Z$ larger than protons. We use the
notation $m$ for the proton mass. The acceleration in old SNRs
($t\gtrsim3\times10^{4}-10^{5}$ yr) when $p_{\max}/mc<10$ are not considered
in the present paper because Eq. (1) is not applied at low Mach numbers, see
Paper I for detail. [Using the test particle approximation for not modified
shock, Drury et al. (\cite{drury03}) found that the spectrum of accelerated 
particles is
somewhat steeper if the diffusion coefficient is increasing with time 
compared
to the case of constant $D$. This effect is not included in our 
consideration.]

The following steady-state equation determines the energy density $W(k)$ 
($k$
is the wave number) of the magnetohydrodynamic turbulence amplified by the
streaming instability in the cosmic-ray precursor upstream of the supernova
shock:
\begin{equation}
u\nabla W(k)=2(\Gamma_{\mathrm{cr}}-\Gamma_{\mathrm{l}}-\Gamma_{\mathrm{nl}
})W(k). \label{2}
\end{equation}
Here the l.h.s. describes the advection of turbulence by highly supersonic 
gas
flow. The terms on the r.h.s. of the equation describe respectively the wave
amplification by cosmic ray streaming, the linear damping of waves in
background plasma, and the nonlinear wave-wave interactions that may limit 
the
amplitude of turbulence. The equation for wave growth rate at the shock
\begin{equation}
\Gamma_{\mathrm{cr}}(k)=\frac {C_{\mathrm{cr}}
(a)\xi_{\mathrm{cr}}u_{\mathrm{sh}}
^{3}k^{1-a}}{\left(  1+A_{\mathrm{tot}}^{2}\right)  ^{(1-a)/2}
cV_{\mathrm{a}}\varphi(p_{\mathrm{\max}})r_{\mathrm{g}0}^{a}}
\label{3}
\end{equation}
was suggested in Paper I as the generalization of equation derived for the
case of weak random field (Berezinskii et al. \cite{berezinsky90}).
Here $V_{\mathrm{a}}$
$=B_{0}/\sqrt{4\pi\rho}$ is the Alfven velocity ($\rho$ is the gas density),
$A=\delta B/B_{0}$ is the dimensionless wave amplitude, and $r_{\mathrm{g}
0}=mc^{2}/eB_{0}$. The ion-neutral and electron-ion collisions usually
determine the linear damping processes in the thermal space plasma. The
Kolmogorov-type nonlinearity with a simplified expression
\begin{equation}
\Gamma_{\mathrm{nl}}=(2C_{\mathrm{K}})^{-3/2}V_{\mathrm{a}}kA(>k)\approx
0.05V_{\mathrm{a}}kA(>k) \label{4}
\end{equation}
at $C_{\mathrm{K}}=3.6$ (as it follows from the numerical simulations by 
Verma
et al. \cite{verma}) was used in Paper I. The wave-particle interaction is 
of resonant
character and the resonance condition is $k_{\mathrm{res}}r_{\mathrm{g}}=\sqrt
{1+A_{\mathrm{tot}}^{2}}$, where the Larmor radius is defined through the
regular field $B_0$ $r_{\mathrm{g}}=pc/ZeB_{0}$ , and $A_{\mathrm{tot}}$ 
is the total 
amplitude of random field. The particle scattering leads to
the cosmic-ray diffusion with the diffusion coefficient 
$D=(1+A_{\mathrm{tot}
}^{2})^{1/2}vr_{\mathrm{g}}\left[  3A^{2}(>k_{\mathrm{res}})\right]  ^{-1}$.

Eq. (2) allows finding the following approximate equation for the
dimensionless amplitude of the total random magnetic field produced by the
cosmic-ray streaming instability at the shock (Paper I):

\[
\frac{3u_{\mathrm{sh}}^{2}A_{\mathrm{tot}}^{2}}{2v(1+A_{\mathrm{tot}}^{2}
)}+\frac{V_{\mathrm{a}}A_{\mathrm{tot}}}{(2C_{\mathrm{K}})^{3/2}}
=
\]
\begin{equation}
\frac{C_{\mathrm{cr}}(a)\xi_{\mathrm{cr}}u_{\mathrm{sh}}^{3}}{cV_{\mathrm{a}
}\varphi(p_{\mathrm{\max}})(p_{\mathrm{\max}}/mc)^{-a}\sqrt{1+A_{\mathrm{tot}
}^{2}}}. \label{5}
\end{equation}
To study the effect of nonlinear interactions, 
the term with linear damping in  Eq.(2) was omitted in Eq.(5);  
$C_{\mathrm{cr}}(a)=27[4(5-a)(2-a)]^{-1}$. The maximum particle 
momentum
satisfies the equation
\begin{equation}
\frac{p_{\mathrm{\max}}}{mc}=\frac{3\varkappa A_{\mathrm{tot}}^{2}
u_{\mathrm{sh}}R_{\mathrm{sh}}}{\sqrt{1+A_{\mathrm{tot}}^{2}}vr_{\mathrm{g}0}
}. \label{6}
\end{equation}

In the high velocity limit, when $u_{\mathrm{sh}}\gg4aC_{\mathrm{cr}}
\xi_{\mathrm{cr}}c[9(2C_{\mathrm{K}})^{3/2}]^{-1}$ and $u_{\mathrm{sh}}
\gg3V_{\mathrm{a}}[2aC_{\mathrm{cr}}\xi_{\mathrm{cr}}]^{-1}$, the advection
term dominates over the nonlinear dissipation term in the l.h.s. of Eq. (5)
and the wave amplitude is large, $A_{\mathrm{tot}}\gg1$. The maximum momentum 
of
accelerated particles and the amplified magnetic field are given then by the
approximate equations
\begin{equation}
\frac{p_{\mathrm{\max}}}{mc}\approx2\varkappa aC_{\mathrm{cr}}\xi
_{\mathrm{cr}}u_{\mathrm{sh}}^{2}R_{\mathrm{sh}}\left(  r_{\mathrm{g}
0}V_{\mathrm{a}}c\right)  ^{-1}, \label{7}
\end{equation}
and

\begin{equation}
A_{\mathrm{tot}}\approx\frac{2u_{\mathrm{sh}}}{3V_a}
aC_{\mathrm{cr}}\xi_{\mathrm{cr}
} \label{8}
\end{equation}

The cosmic ray diffusion coefficient depends on particle Larmor
radius as $D\propto vr_{\mathrm{g}}^{1-a}$ at $p\leq p_{\max}$ in this case.

In the low velocity limit, when $u_{\mathrm{sh}}\ll\left[  4V_{\mathrm{a}}
^{3}c^{2}\left(  \pi aC_{\mathrm{cr}}(2C_{\mathrm{K}}\right)  ^{3}
\xi_{\mathrm{cr}})^{-1}\right]  ^{1/5}$ and $u_{\mathrm{sh}}\ll\left[
V_{\mathrm{a}}^{2}c\left(  aC_{\mathrm{cr}}(2C_{\mathrm{K}})^{3/2}
\xi_{\mathrm{cr}}\right)  ^{-1}\right]  ^{1/3}$, the nonlinear dissipation
term dominates over the advection term in the l.h.s. of Eq. (5) and the wave
amplitude is small, $A_{\mathrm{tot}}\ll1$. The maximum momentum of 
accelerated
particles and the amplified magnetic field are given then by the approximate
equations
\begin{equation}
\frac{p_{\max}}{mc}\approx24\varkappa a^{2}C_{\mathrm{cr}}^{2}C_{\mathrm{K}
}^{3}\xi_{\mathrm{cr}}^{2}u_{\mathrm{sh}}^{7}R_{\mathrm{sh}}\left(
r_{\mathrm{g}0}V_{\mathrm{a}}^{4}c^{3}\right)  ^{-1}, \label{9}
\end{equation}
and
\begin{equation}
A_{\mathrm{tot}}\approx aC_{\mathrm{cr}}(2C_{\mathrm{K}})^{3/2}\xi_{\mathrm{cr}
}u_{\mathrm{sh}}^{3}\left(  cV_{\mathrm{a}}^{2}\right)  ^{-1} 
\label{10}
\end{equation}
assuming that this value of amplified field exceeds the value of random
interstellar magnetic field. The cosmic ray diffusion coefficient depends on
particle Larmor radius as $D\propto vr_{\mathrm{g}}^{1-2a}$ at $p\leq 
p_{\max
}$. (Notice the misprint in the numerical coefficient in the first equality
of Eq. (19) in Paper I that is analogous to the present Eq. (9).)

%figure1

\begin{figure*}[tbp]
\centering
\includegraphics[width=10.0cm]{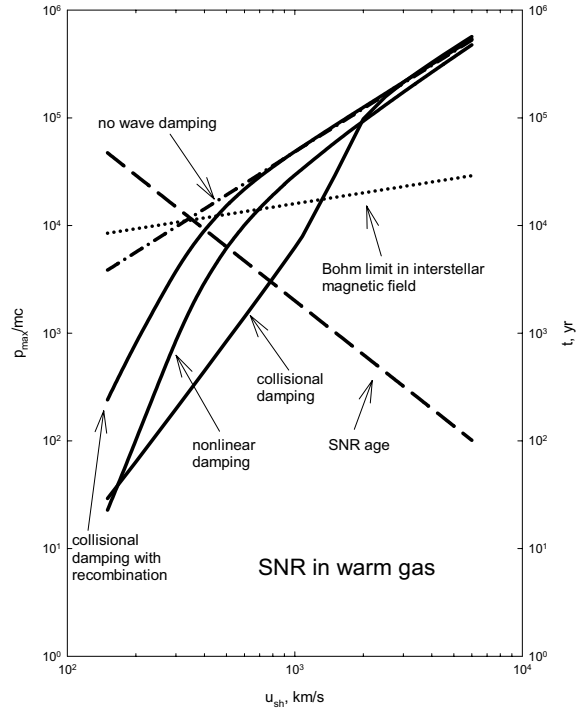}
\caption{The maximum momentum of accelerated protons $p_{\max}$ in units 
$mc$
as a function of shock velocity $u_{\mathrm{sh}}$ at the Sedov stage of 
$\sup
$ernova remnant evolution in warm interstellar gas. Three solid lines
correspond to three cases of wave dissipation considered separately: 
nonlinear
wave interactions; damping by ion-neutral collisions at constant density of
neutral atoms; damping by ion-neutral collisions when the diffuse neutral 
gas
restores its density after complete ionization by the radiation from 
supernova
explosion. The dashed line presents the age of a supernova remnant $t$ 
(plotted
on the right ordinate) as a function of shock velocity. The dotted line 
shows
the Bohm limit on maximum particle momentum calculated for the interstellar
magnetic field strength. The dash-dot line gives the maximum particle 
momentum
when the wave dissipation is not taken into account.}
\label{Fig1}
\end{figure*}

Fig.1 illustrates the results of calculations of $p_{\max}$ at the Sedov 
stage
of SNR evolution at $\mathcal{E}=10^{51}$ erg in the warm interstellar gas
with the temperature $T=8\times10^{3}$ K, the average density $n_{0}=0.4$
cm$^{-3}$ that includes small interstellar clouds, the intercloud density
$n=0.1$ cm$^{-3}$, the number density of ions $n_{\mathrm{i}}=0.03$ 
cm$^{-3}$,
the interstellar magnetic field value $B_{0}=5$ $\mu$G, see Paper I. The 
time
dependence of the shock radius and the shock velocity are given by the 
following
equations (the Sedov solution, see e.g. Ostriker \& McKee \cite{ostriker}):
\[
R_{\mathrm{sh}}=4.3\left(  \mathcal{E}_{51}/n_{0}\right)  ^{1/5}
t_{\mathrm{Kyr}}^{2/5}\;{\mathrm{pc, }}
\]
\begin{equation}
u_{\mathrm{sh}}=1.7\times
10^{3}\left(  \mathcal{E}_{51}/n_{0}\right)  ^{1/5}t_{\mathrm{Kyr}}
^{-3/5}\; {\mathrm{km/s,}} \label{11}
\end{equation}
where we assume that the ultrarelativistic gas of cosmic rays mainly
determines the pressure behind the shock. The value $\varkappa=0.04$ was
assumed in the calculations in Fig. 1. Three solid lines correspond to the
three cases of wave dissipation considered separately: the nonlinear wave
interactions; the damping by ion-neutral collisions at constant gas density;
the damping by ion-neutral collisions when the diffuse neutral gas restores
its density after complete ionization by the radiation from the SN burst. 
For
the last two curves, the dissipation of wave due to the ion-neutral 
collisions
with damping rate
\begin{equation}
\Gamma_{\mathrm{l}}=\frac{\nu_{\mathrm{in}}}{2}\left(  1+\left(
1+A_{\mathrm{tot}}^{2}\right)  ^{-1}\left(  (1+\frac{n_{\mathrm{i}}
}{n_{\mathrm{H}}})\frac{\nu_{\mathrm{in}}}{kV_{\mathrm{a}}}\right)
^{2}\right)  ^{-1} \label{12}
\end{equation}
was taken into account whereas the term $\Gamma_{\mathrm{nl}}$ that 
describes
the nonlinear dissipation was omitted. Here $\nu_{\mathrm{in}}=n_{\mathrm{H}
}\left\langle v_{\mathrm{th}}\sigma\right\rangle \approx8.4\times
10^{-9}(T/10^{4}$ \textrm{K}$)^{0.4}(n_{\mathrm{H}}/1$ \textrm{cm}$^{-3})$
s$^{-1}$ for the temperature $T\sim10^{2}-10^{5}$ K is the frequency of
ion-neutral collisions with the cross section $\sigma$ averaged over 
velocity
distribution of thermal particles, $n_{\mathrm{H}}$ is the number density of
neutral hydrogen. (Notice some correction of the ``collisional damping'' 
curve in Fig.1 comparing to corresponding Figure 2 in paper I.)
The maximum
energy of protons accelerated by SN shocks at the early Sedov stage is close
to $3\times10^{14}$ eV that exceeds the Bohm limit calculated for the
interstellar magnetic field value by one order of magnitude. The maximum
energy decreases to about $10^{10}$ eV at the end of the Sedov stage that is
much less than the Bohm limit calculated for the interstellar magnetic field
value. In particular, the particle energy is less than $10^{13}$ eV at
$t>3\times10^{3}$ yr and this may explain the absence of a TeV $\gamma$-ray
signal from many SNRs (Buckley et al. \cite{buckley}, Aharonian et al. 
\cite{aharonian})
where the gamma-rays could in
principle be produced through $\pi^{0}$ decays if sufficiently energetic
cosmic rays were present.

With the extreme choice of parameters of the flying apart young SNR 
envelope,
it was found (Bell \& Lucek \cite{bell01}, Paper I) that the maximum 
particle energy
may reach the ultra high energies. The estimate of the highest particle 
energy
according to Paper I is $E_{\max}\approx2\times10^{17}Z(u_{\mathrm{sh}
}/3\times10^{4}$ \textrm{km} \textrm{s}$^{-1})^{2}(\varkappa/0.1)\xi
_{\mathrm{cr}}M_{\mathrm{ej}}^{1/3}n^{1/6}$ eV at the end of a free 
expansion
stage which precedes the Sedov stage (here $M_{\mathrm{ej}}$ is the mass of
ejecta in solar masses). We shall see below that this promising estimate is 
in
some sense devaluated by the results of calculations of particle flux - the
flux turns out to be low at the highest energies which can be achieved in 
the process
of acceleration.

\section{Evolution of SNR Shocks}

The typical source of galactic cosmic rays is most probably associated with
the core collapse supernova, Type II SNe, that is the final stage of 
evolution
for stars more massive than about $8$ solar masses while on the main 
sequence.
The massive star before the explosion goes through the sequence O-star 
stage,
Red Super Giant star stage, and through the
Wolf-Rayet stage for the most massive progenitors ($>20M_{\odot}$) that
give the rare in occurrence Type Ib/c SNe.
The fast wind of a massive progenitor star on the main
sequence produces a big bubble of hot rarefied gas with the temperature 
about
$10^{6}$ K in the surrounding interstellar medium, see Weaver et al. 
(\cite{weaver}),
Lozinskaya (\cite{lozinskaya}). The typical Type II SN goes through the Red 
Super Giant
phase before the explosion and this process is accompanied by the flow of a
low-velocity dense wind. Thus, immediately after the supernova burst, the
shock propagates through the wind of a Red Super Giant star then through the
hot bubble and finally it enters the interstellar medium. Our
calculations will be done for the ejecta mass 
$M_{_{\mathrm{ej}}}=1M_{\odot}$
(the solar mass). The spherically symmetric distribution of gas density in 
the
stellar wind is $n_{\mathrm{w}}=\dot{M}/(4\pi m_{\mathrm{a}
}u_{\mathrm{w}}r^{2})$, where $\dot{M}=10^{-5}\dot{M}_{-5}$
(solar mass)/yr is the mass loss rate, $m_{\mathrm{a}}=1.4m$ is the
mean interstellar atom mass per hydrogen nucleus, the wind velocity
$u_{\mathrm{w}}=10^{6}u_{\mathrm{w},6}$ cm/s. The stellar wind magnetic 
field
has the shape of the Parker spiral similar to the case of interplanetary
magnetic field (Parker, 1958). At relatively large distances from the 
surface
of the star that are of interest here, the magnetic field has predominately
azimuthal structure and its value is $B_{0}=B_{\ast}r_{\ast}^{2}\Omega
\sin\theta/(u_{\mathrm{w}}r)$ where $B_{\ast}$ is the surface magnetic field
strength at the star radius $r_{\ast}$, $\Omega$ is the angular velocity of
star rotation, and $\theta$ is the polar angle. Hence $B_{0}(r)r=2\times
10^{13}u_{\mathrm{w},6}^{-1}\sin\theta$ G$\times$cm at $B_{\ast}=1$ G,
$r_{\ast}=3\times10^{13}$ cm, $\Omega=3\times10^{-8}$ s$^{-1}$ that gives
$B_{0}\approx6$ $\mu$G at the distance $r=1$ pc from the star.

Below we shall also use the following set of parameters of the medium
surrounding the Type II SN: the radius of spherical Red Super Giant wind
$R_{\mathrm{w}}=2$ pc, the star mass loss $\dot{M}_{-5}=1$, and the wind 
velocity
$u_{\mathrm{w},6}=1$. The radius of the spherical bubble of hot gas 
$R_{b}=60$
pc , the gas density in the bubble $n_{\mathrm{b}}=1.5\times10^{-2}$ 
cm$^{-3}
$, the magnetic field there $B_{\mathrm{b}}=5$ $\mu$G. The gas density in 
the
undisturbed interstellar medium around the bubble is assumed to be equal to
$n_{0}=1$ cm$^{-3}$ (physically, the value of $n_{0}$
determines$\ n_{\mathrm{b}}$, see Weaver et al. \cite{weaver}). The hot 
bubble is
separated by the dense thin shell from the interstellar gas. The accepted
parameters are close to those assumed by Berezhko \& V\"{o}lk 
(\cite{berezhko00})
in their
analysis of gamma-ray production in SNRs. The lengthy discussion and the
additional references can be found there.

The considerable fraction of cosmic rays is probably accelerated in Type Ia
SNe (their explosion rate in the Galaxy is about $1/4$ of supernovae Type
II rate). These supernovae are caused by the thermonuclear explosions of
compact white dwarfs following mass accretion. The characteristic mass of a
progenitor star and the mass of ejecta are $1.4$ solar mass. The progenitor
stars do not appear to have observable amount of mass loss nor do they emit
ionizing radiation that could essentially modify the ambient medium around 
the
star. We assume that the SNR shock goes through the uniform weakly ionized
interstellar medium with density $1$ cm$^{-3}$, and the magnetic field $7$
$\mu$G.

The two asymptotic regimes of the propagation of SNR shock - the ejecta
dominated stage and the adiabatic stage - are instructive to consider.

The adiabatic regime was mentioned earlier, see the Sedov solution (11) for
the shock moving in the gas with constant density, and it refers to the 
stage
of SNR evolution when the mass of swept-up gas significantly exceeds the 
mass
of ejecta. This condition is fulfilled in the case of the medium with 
constant density
at $R_{\mathrm{sh}}>R_{0}=(3M_{\mathrm{ej}}/4\pi m_{\mathrm{a}}n_{0}
)^{1/3}=1.9(M_{\mathrm{ej}}/M_{\odot}n_{0})^{1/3}$ pc, $t_{0}>R/u_{0}
\approx190n_{0}^{-1/3}$ yr, where $u_{0}\sim10^{9}$ cm/s is the ejecta 
initial
velocity. The adiabatic regime for the SNR shock moving through the 
progenitor
star wind is described by the equations (at $u_{\mathrm{sh}}\gg 
u_{\mathrm{w}
})$:
$$
R_{\mathrm{sh}}=7.9\left(  \frac{\mathcal{E}_{51}u_{\mathrm{w},6}}
{\dot{M}_{-5}}\right)  ^{1/3}t_{\mathrm{Kyr}}^{2/3}
\; {\mathrm{pc, }}
$$
\begin{equation}
u_{\mathrm{sh}}=5.2\times10^{3}\left(  \frac
{\mathcal{E}_{51}u_{\mathrm{w},6}}{\dot{M}_{-5}}\right)
^{1/3}t_{\mathrm{Kyr}}^{-1/3}\; {\mathrm{km/s,}}\label{13}
\end{equation}
see Ostriker \& McKee (1998). As in Eq. (11), we assume that the
ultrarelativistic gas of cosmic rays mainly determines the pressure behind 
the
shock. Eq. (13) is valid when the mass of swept-up gas is relatively large 
and
$R_{\mathrm{sh}}>R_{0}=M_{ej}u_{\mathrm{w}}/\dot{M}\approx
1(M_{\mathrm{ej}}/M_{\odot})u_{\mathrm{w},6}/\dot{M}_{-5}$pc.

The quantity $\rho u_{\mathrm{sh}}^{2}R_{\mathrm{sh}}^{3}=K\mathcal{E}$ is
conserved for the considered adiabatic shocks. The constant $K\approx0.16$ 
for
the solution (11), and $K\approx0.34$ \ for the solution (13). In the 
general
case of the power-law gas distribution $\rho=\rho_{0}(r)r^{-s}$, $s<5$, the
adiabatic shock evolution is described by the equations $R_{\mathrm{sh}}
=(\eta(s)\mathcal{E}/\rho_{\mathrm{0}})^{\frac{1}{5-s}}t^{\frac{2}{5-s}}$, 
and
$u_{sh}=\frac{2}{5-s}(\eta(s)\mathcal{E}/\rho_{\mathrm{0}})^{\frac{1}{5-s}
}t^{-\frac{3-s}{5-s}}$, where $\eta$ is constant at fixed $s$, that gives 
the
general formula $K=\frac{4\eta(s)}{(5-s)^{2}}$ ( the values of $\eta(s)$ 
were
given by Ostriker \& McKee, \cite{ostriker}).

The ejecta dominated stage precedes the adiabatic one. As long as the mass 
of
the ejecta is large compared to the swept-up mass, the blast wave is moving
with relatively weak deceleration. At this stage shortly after the 
explosion,
the structure of the flying apart envelope of the presupernova star is
important for the shock evolution. Actually, the blast wave consists of two
shocks, the forward shock and the reverse shock, with the contact
discontinuity surface between them. This surface separates the shocked wind 
or
interstellar gas downstream of the forward shock from the shocked envelope 
gas
that fills the downstream region of the reverse shock. The reverse shock 
lags
behind the forward shock and enters the dense internal part of the flying
apart star by the time of the beginning of Sedov stage. Though it can not be
well justified approximation for the very young SNRs, we ignore below the
cosmic ray acceleration at the reverse shock compared to the forward shock 
and
use the notation $R_{\mathrm{sh}}$ for the radius of forward shock 
(Berezinsky
\& Ptuskin (\cite{berezinsky89}) considered the cosmic ray acceleration by 
both shocks, see
also Yoshida \& Yanagita (\cite{yoshida})). The outer part of the star that 
freely
expands after the SN explosion has a power law density profile $\rho
_{\mathrm{s}}\varpropto r^{-k}$, see e.g. Chevalier \& Liang 
(\cite{chevalier89}).
The value
of $k$ typically lies between $6$ and $14$. The value $k\approx7$ is
characteristic of the SNe Type Ia, and $k\approx10$ is typical of the SNe 
Type
II. The self similar solution for the blast wave at the ejecta dominated 
stage
was found by Chevalier (\cite{chevalier82}) and Nadyozhin 
(\cite{nadyozhin81},
\cite{nadyozhin85}).
It was shown that at the
age larger than about one week, the evolution of the shock at the ejecta
dominated stage can be approximately described by the power-law dependence
$R_{\mathrm{sh}}\propto t^{\lambda}$ where the expansion parameter
$\lambda=\frac{k-3}{k}$ for the explosion in the uniform medium, and
$\lambda=\frac{k-3}{k-2}$ for the explosion in the wind of a presupernova 
star
(for $k>5$ ejecta). In particular, using the results of two mentioned above 
papers,
one can obtain the following equations
\[
R_{{\mathrm{sh}}}=5.3\left(  \frac{\mathcal{E}_{51}^{2}M_{\odot}}
{n_{0}M_{\mathrm{ej}}}\right)  ^{1/7}t_{\mathrm{Kyr}}^{4/7}\; {\mathrm{pc,
}}
\]
\begin{equation}
u_{\mathrm{sh}}=2.7\times10^{3}\left(  \frac{\mathcal{E}_{51}^{2}M_{\odot}
}{n_{0}M_{\mathrm{ej}}}\right)  ^{1/7}t_{\mathrm{Kyr}}^{-3/7}\;\mathrm{km/s}
\label{14}
\end{equation}
for the Type Ia SN explosion in the uniform interstellar medium at $k=7$;
$$
R_{_{\mathrm{sh}}}=7.7\left(  \frac{\mathcal{E}_{51}^{7/2}u_{\mathrm{w}
,6}M_{\odot}^{5/2}}{\dot{M}_{-5}M_{\mathrm{ej}}^{5/2}}\right)
^{1/8}t_{\mathrm{Kyr}}^{7/8}\;\mathrm{pc,\;}
$$
\begin{equation}
u_{\mathrm{sh}}=6.6\times
10^{3}\left(  \frac{\mathcal{E}_{51}^{7/2}u_{\mathrm{w},6}M_{\odot}^{5/2}
}{\dot{M}_{-5}M_{\mathrm{ej}}^{5/2}}\right)  ^{1/8}t_{\mathrm{Kyr}
}^{-1/8}\;\mathrm{km/s}\label{15}
\end{equation}
for the Type II SN explosion in the wind of a presupernova star at $k=10$.

Following the approach of Truelove \& McKee (\cite{truelove}), one can 
describe the shock
produced by the Type Ia SN using the continuous solution which coincides 
with
the ejecta dominated equation (14) until the moment 
$t_{0}=260(M_{\mathrm{ej}
}/1.4M_{\odot})^{5/6}\mathcal{E}_{51}^{-1/2}n_{0}^{-1/3}$ yr, and is given 
by
the equations
$$
R_{\mathrm{sh}}=4.3\left(  \mathcal{E}_{51}/n_{0}\right)  ^{1/5}
t_{\mathrm{Kyr}}^{2/5}\times
$$
\begin{equation}
\left(  1-\frac{0.06(M_{\mathrm{ej}}/M_{\odot})^{5/6}
}{\mathcal{E}_{51}^{1/2}n_{0}^{1/3}t_{\mathrm{Kyr}}}\right)  ^{2/5}
\; {\mathrm{pc, \ }}\label{16}
\end{equation}
\[
u_{\mathrm{sh}}=1.7\times10^{3}\left(  \mathcal{E}_{51}/n_{0}\right)
^{1/5}t_{\mathrm{Kyr}}^{-3/5}\times
\]
\[
\left(  1-\frac{0.06(M_{\mathrm{ej}}/M_{\odot
})^{5/6}}{\mathcal{E}_{51}^{1/2}n_{0}^{1/3}t_{\mathrm{Kyr}}}\right)
^{-3/5}\; {\mathrm{km/s}}
\]
at a later time $t>t_{0}$. It is evident from Eq. (16) that the adiabatic
asymptotics (11) holds at $t\gg t_{0}$.

The evolution of the Type II SN shock first follows the ejecta dominated
solution (15) in a presupernova wind and then, while still moving in the 
wind,
it enters the adiabatic regime at the distance $r\sim1$ pc. The subsequent
evolution proceeds in the medium with a complicated structure described 
above
for the Type II SN. The fairly accurate solution for the SNR evolution 
during
this period can be obtain in the ''thin-shell'' approximation, e.g. Ostriker
\& McKee (\cite{ostriker}), Bisnovatyi-Kogan \& Silich (\cite{kogan}).
Using this
approximation for the strong shock and assuming the spherically symmetric
distribution of the circumstellar gas density $\rho(r)$, we come to the
following equations where the shock velocity $u_{\mathrm{sh}}$ and the SNR 
age
$t$ are parameterized as functions of the shock radius $R_{\mathrm{sh}}$ 
(see
Appendix for the derivation of these equations):
\[
u_{\mathrm{sh}}(R_{\mathrm{sh}})=\frac{\gamma_{\mathrm{ad}}+1}{2}\left[
\frac{12(\gamma_{\mathrm{ad}}-1)\mathcal{E}}{(\gamma_{\mathrm{ad}}
+1)M^{2}(R_{\mathrm{sh}})R_{\mathrm{sh}}^{6(\gamma_{\mathrm{ad}}
-1)/(\gamma_{\mathrm{ad}}+1)}}\cdot \right.
\]
\begin{equation}
\left. \cdot \int_{0}^{R\mathrm{sh}}drr^{6\left(
\frac{\gamma_{\mathrm{ad}}-1}{\gamma_{\mathrm{ad}}+1}\right)  -1}M(r)\right]
^{1/2},\; \label{17}
\end{equation}
\[
t(R_{\mathrm{sh}})=\int_{0}^{R_{\mathrm{sh}}}\frac{dr}{u_{\mathrm{sh}}(r)},
\]
where $\gamma_{\mathrm{ad}}$ is the adiabatic index ($\gamma_{\mathrm{ad}
}=4/3$ if the pressure downstream of the shock is determined by the
relativistic particles), $M(R)=M_{ej}+4\pi$ $\int_{0}^{R}drr^{2}\rho(r)$ is
the mass of the swept up gas. The self similar solution by Chevalier and
Nadyozhin is not explicitly reproduced by Eqs. (17). The solutions (15) and
(17) are fitted together at the transition from the ejecta dominated regime 
to
the adiabatic regime (at $r\sim0.3$ pc) in our numerical simulations of 
cosmic ray acceleration in the
Type II SNRs described below.

It is worth noting that the energy loss of SNR in a form of escaping cosmic
rays is not taken into account in the solutions for shock evolution that 
were
described in this Sections. In fact the shock evolution is only
approximately adiabatic.

\section{Average Spectrum of Cosmic Rays Injected in the Interstellar 
medium}

At a given SNR age $t$, the cosmic rays are accelerated up to maximum 
momentum
$p_{\max}(t)$. Also, particles with $p>p_{\max}(t)$ cannot be confined in 
the
precursor of the shock even if they were accelerated earlier. Thus particles
accelerated to the maximum energy escape from a SNR
(see also Berezhko \& Krymsky 1988). Let us
estimate the flux of these run-away particles. We consider the simplified
approach for maximum energy of accelerated particles and take the dependence
of diffusion on momentum in the following simplified form:
\[
D(p)=D_{0}<<Ru_{\mathrm{sh}},\;p\leqslant p_{\mathrm{max}}(t),
\]
\begin{equation}
D(p)=D_{\mathrm{m}}>>Ru_{\mathrm{sh}},\;p>p_{\mathrm{max}}(t). \label{18}
\end{equation}

The spectrum of accelerated particles in this case has a very steep cut-off 
at
$p>p_{\mathrm{max}}$ (cf. Eq (1)) and the spectrum of run-away particles
beyond $p_{\mathrm{max}}$ can be approximated by $\delta$-function. To find 
the
equation for these particles, let us integrate the equation for cosmic-ray
distribution function
\begin{equation}
\frac{\partial f}{\partial t}-\nabla D\nabla f+\mathbf{u}\nabla f-\frac
{\nabla\mathbf{u}}{3}p\frac{\partial f}{\partial p}=0 \label{19}
\end{equation}
on momentum $p$ from $p_{\mathrm{max}}$ to $p_{\mathrm{max}}+\Delta p$, 
where
$\Delta p<<p_{\mathrm{max}}$ and larger then the width of run-away particle
spectrum. Denoting $G=\int_{p_{\mathrm{max}}}^{p_{max}+\Delta p}fdp$ one
obtains from Eq (19):
\[
\frac{\partial G}{\partial t}+\mathbf{u}\nabla G=
\]
\begin{equation}
\nabla D_{\mathrm{m}}\nabla G-\frac{\partial p_{\mathrm{max}}}{\partial
t}f(p_{\mathrm{max}}-0)-\frac{\nabla\mathbf{u}}{3}p_{\mathrm{max}
}f(p_{\mathrm{max}}-0). \label{20}
\end{equation}
Since diffusion coefficient of run-away particles is large, the advection
terms play no role in this equation and the last two terms can be considered
as source of particles. The total source of run-away particles is given by 
the
volume integral of this terms. As a result, the source spectrum of run-away
particles has the form
\[
q(p,t)=-\delta(p-p_{\mathrm{max}})\times
\]
\begin{equation}
\int d^{3}r\left(  \frac{\partial p_{\mathrm{max}}}{\partial t}+\frac
{\nabla\mathbf{u}}{3}p_{\mathrm{max}}\right)  f(p_{\mathrm{max}}
-0,\mathbf{r}). \label{21}
\end{equation}
The integration here is performed over the domain where the integrand is
negative. The integral $4\pi\int dpp^{2}q(p,t)$ has dimensions number of
particles per unit time.

Below we consider the case of spherically symmetric SN shock with linear
velocity profile at $r<R_{\mathrm{sh}}$:
\begin{equation}
u=\left(  1-\frac{1}{\sigma}\right)  u_{\mathrm{sh}}(t)r/R_{\mathrm{sh}}(t),
\label{22}
\end{equation}
where $\sigma $ is the total shock compression ratio. It includes a thermal 
subshock and
a cosmic ray precursor. Linear profile of velocity (22) is a good 
approximation of Sedov's solution and it can be considered as a very 
approximate one at the ejecta-dominated stage.
Since the shock is partially modified in the 
presence of cosmic rays,
we should not assume
any relation between the shock compression ratio $\sigma $ and the spectral 
index of accelerated
particles $4-a$ (recall that $4-a=3\sigma/(\sigma-1)$ for not modified 
shocks). We accept the value $\sigma=7$
in our calculations.
The preshock at $r>R_{\mathrm{sh}}$ is created by the cosmic
ray pressure gradient. Its width is small in comparison with the shock 
radius under the conditions
given by Eq. (18) and the plane shock approximation can be used. Since the 
cosmic ray
pressure dominates the gas pressure in the precursor region, its gradient is 
proportional to
the velocity gradient
$\partial P_\mathrm{cr}/ \partial r=\rho u_\mathrm{sh}\partial u/\partial 
r$, where
$\rho $ is the circumstellar medium density. We also use an assumption that 
cosmic
ray pressure at the shock is some fraction $\xi _{\mathrm{cr}}$ of the 
upstream
momentum flux, see Eq. (1).
Now assuming that $f(p_{\mathrm{max}})$ is proportional to the cosmic ray 
pressure
the expression (21) for the run-away particle source takes the form
\[
q(p,t)=4\pi\delta(p-p_{\mathrm{max}})\left(  \frac 13
\left( 1-\frac 1\sigma -\frac {\xi_{\mathrm{cr}}}2\right)
R^{2}u_{\mathrm{sh}} pf_{0}(p)-\right.
\]
\begin{equation}
\left.  \int_{0}^{R}r^{2}drf(p_{\mathrm{max}},r)\left(  \frac{\partial
p_{\mathrm{max}}}{\partial t}+\frac{\sigma -1}{\sigma }p_{\mathrm{max}}\frac
{u_{\mathrm{sh}}}{R}\right)  \right)  \label{23}
\end{equation}
%where $\gamma=3\sigma/(\sigma-1)$ and $f_{0}(p)$
%is the distribution function at the shock front.
The first term in this expression describes the particles
which runs away from the shock front, and the second term describes the
particles escaping from the shock interior. 
In principle, the turbulence downstream the strong shock might be 
enhanced that would result in the small cosmic ray diffusion coefficient. In 
this case the particles do not run-away from the 
downstream and the second term 
in Eq. (23) should be omitted. If the turbulence downstream is maintained 
by the same process of the cosmic ray streaming instability as in the 
upstream region, 
the downstream diffusion coefficient is comparable to the upstream diffusion 
coefficient for particles with $p\sim p_{max}$. We shall furhter assume 
that particles can run away both from upstream and downstream of the shock. 
The uncertainty of the efficiency of run-away process in the inner part of 
SNR does not qualitatively change the conclusion about the average source 
spectrum of cosmic rays calculated later in this Section and shown in Fig.2.

The distribution function of 
particles
with $p\leqslant p_{\mathrm{max}}$ can be found using the solution of 
transport
equation (19) at $r<R_{\mathrm{sh}}$ with the boundary condition
$f(p,r=R_{\mathrm{sh}},t)=f_{0}$ by the method of characteristics. As the
result
\[
q(p,t)=4\pi\delta(p-p_{\mathrm{max}})\left[
\frac 13
\left( 1-\frac 1\sigma -\frac {\xi_{\mathrm{cr}}}2\right)
R^{2}u_{\mathrm{sh}} pf_{0}(p)
+\right.
\]
\[
\left(  -\frac{\partial p_{\mathrm{max}}}{\partial t}-\frac{\sigma 
-1}{\sigma
}p_{\mathrm{max}}\frac{u_{\mathrm{sh}}}{R}\right)  
\int_{0}^{t}\frac{dt}{\sigma}^{\prime}R^{2}
(t^{\prime})u_{\mathrm{sh}}(t^{\prime})\times
\]
\begin{equation}
\left. f_{0}\left(  p\left(  \frac{R(t)}{R(t^{\prime})}\right)
^{1-\frac 1\sigma },t^{\prime}\right)  \left(  
\frac{R(t)}{R(t^{\prime})}\right)
^{3-\frac 3\sigma }\right]. \label{24}
\end{equation}
The expression in brackets in front of the integral in Eq. (24) should be
positive that means that the particles lose energy adiabatically slower then
the maximum energy decreases. For the opposite sign, the adiabatic losses of
particles are faster then the decrease of maximum energy and the particles
don't run away from downstream of the shock. They can run away at later time
if at that time the decrease of maximum momentum will be faster.

The average source power $Q(p)$\ of run-away cosmic rays per unit volume in
the galactic disk is obtained by the integrating $q$ with respect of $t$ \ 
and
by the averaging over many SN explosions: $Q(p)=\nu_{\mathrm{sn}}\int_{\min
}^{\max}dtq(p,t)$, where $\nu_{\mathrm{sn}}$ is the average frequency of SN
explosions per unit volume of the galactic disk. Changing the variable of
integration from $t$ to $R_{\mathrm{sh}}$ ($dR_{\mathrm{sh}}=u_{\mathrm{sh}
}dt$) one can derive the following equation:
\[
Q(p)=\frac{3a\xi_{\mathrm{cr}}\nu_{\mathrm{sn}}}{cp^{4}
\left| \frac{d\ln(p_{\max})}{d\ln
(R)}\right|
}\left[  \frac
{\left( 1-\frac 1\sigma -\frac {\xi_{\mathrm{cr}}}2\right)
\rho (R)u^2_{\mathrm{sh}}(R) R^{3}}{
3\left(  1-\frac{1}{1+a}\left(  \frac{mc}{p_{\max}(R)}\right)  ^{a}\right)
}+\right.
\]
\[
\left( \frac 1\sigma -1 -\frac{d\ln(p_{\max})}{d\ln(R)} \right)
\int_{R_{\min}}^{R}\frac{dR^{\prime}}{\sigma}\frac{\rho(R^{\prime})u_{sh}
^{2}(R^{\prime})R^{\prime2}}{\left(  1-\frac{1}{1+a}\left( 
\frac{mc}{p_{\max}
(R')}\right) ^{a}\right) }\times
\]
\[
H\left( p_{\max}(R')-p\left(  \frac{R}{R'}\right) ^{1-\frac 1\sigma }\right) 
\times
\]
\begin{equation}
\left. \left(  \frac{p}{p_{\max}(R^{\prime})}\right)  ^{a}
\left(
\frac{R}{R^{\prime}}\right)  ^{(a-1)(\sigma -1)/\sigma }\right]
_{R=R_{\mathrm{m}}(p)} \label{25}
\end{equation}
Here Eq. (1) for $f_{0}$ with the approximation $\varphi(p)\approx
a^{-1}(p/mc)^{a}-a^{-1}(1+a)^{-1}$ is used, and the condition $\xi
_{\mathrm{cr}}=const$ is assumed. The function $R_{\mathrm{m}}(p)$ in Eq. 
(25)
is defined by the equation $p_{\max}(R_{\mathrm{sh}}=R_{\mathrm{m}}(p))=p$. 
If
the last equation has multiple solutions at given $p$, the summation on all
these solutions should be performed in (25). The physical meaning of
$R_{\mathrm{m}}(p)$ is that it is the value of shock radius when the maximum
energy of accelerating particles is equal to $p$.
%The first term in square
%brackets in the r.h.s. of Eq. (25) should be omitted if $dp_{\max
%}/dR_{\mathrm{sh}}>0$, and
The second term in the r.h.s. of Eq. (25) should be omitted if the 
expression
in round parenthesis in front of the integral is negative.

Let us assume that the maximum momentum is a power law function of the shock
radius, $p_{\max}\propto$ $R^{-\delta}$, the particles are 
ultrarelativistic,
$p>>mc$, and the compression ratio is constant, $\sigma=const$. The 
remarkable
feature of Eq. (25) is then that the expression in square brackets does not
depend on momentum at the adiabatic stage of SNR shock propagation in the
medium with a power-law distribution of gas density, because $\rho
u_{\mathrm{sh}}^{2}R_{\mathrm{sh}}^{3}=K\mathcal{E}$, $K=const$ in this 
case,
see Section 3. The average cosmic ray source power is now given by the 
simple
equation
\[
Q(p)=\frac{3Ka\xi_{\mathrm{cr}}\nu_{\mathrm{sn}}\mathcal{E}}{cp^{4}}\left[
\frac{1}{3\delta}\left( 1-\frac 1\sigma -\frac 
{\xi_{\mathrm{cr}}}2\right)+\right.
\]
\begin{equation}
\left. \frac{1}{\sigma}\frac{1-\frac{\sigma -1}{\sigma \delta}}
{1-\frac{1}{\sigma }+a\left( \delta-1+\frac{1}{\sigma }\right)  }\right]  .
\label{26}
\end{equation}
Here the factors $\delta$, and $(1-\frac{\sigma -1}{\sigma \delta})$ should 
be
positive. The first term in the square brackets describes the particles 
which
run-away from the shock and the second term describes the particles which
run-away from the SNR interior. Consequently, while in the adiabatic regime,
the SNR shock during its evolution produces the run-away particles with the
universal power-law overall spectrum $Q(p)\propto p^{-4}$, whereas the
instantaneous spectrum at the shock is more flat and not universal (see Eq.
(1)) and the instantaneous spectrum of run-away particles has a 
delta-function
form (see Eq. (24)).

The total source power of ultrarelativistic particles calculated with the 
use
of Eq. (26) is $W=4\pi c\int dpp^{3}Q(p)=C\xi_{\mathrm{cr}}\nu_{\mathrm{sn}
}\mathcal{E}\ln(p_{\mathrm{\max}2}/p_{\mathrm{\max}1})$, where
$C=12\pi
Ka\left( \frac{1}{3\delta}\left( 1-\frac 1\sigma -\frac 
{\xi_{\mathrm{cr}}}2\right) +
\frac{1}{\sigma}\frac{1-\frac{\sigma -1}{\sigma \delta}}
{1-\frac{1}{\sigma }+a\left( \delta-1+\frac{1}{\sigma }\right)  } \right)
$;
$\ p_{\mathrm{\max}2}$ and $p_{\mathrm{\max}1}$ are the maximum momenta of
accelerated particles at the beginning and at the end of the adiabatic stage
respectively, thus typically $\ln(p_{\mathrm{\max}2}/p_{\mathrm{\max}
1})\approx10$. It leads to the estimate 
$W\approx0.5(\xi_{\mathrm{cr}}/0.5)\nu
_{\mathrm{sn}}\mathcal{E}$\ for the shock moving in the uniform interstellar
medium. Hence the considerable part of the total available mechanical energy
of SN explosion $\nu_{\mathrm{sn}}\mathcal{E}$\ goes to cosmic rays at
$\xi_{\mathrm{cr}}\sim0.5$. As it is well known, the source spectrum 
$\propto$ $p^{-4}$
or somewhat more steep, and the efficiency of cosmic ray acceleration at the
level $10-30 \% $   are needed to fit the cosmic ray data below the knee in 
the
cosmic ray spectrum at about $4\times10^{15}$ eV in the empirical model
of cosmic ray origin (e.g. Ptuskin (2001), see also discussion below).

At the ejecta dominated stage which precedes the adiabatic stage, the 
average
spectrum of the run-away particles is different from $p^{-4}$. Let us 
consider
the general case and assume that $\rho\propto r^{-s}$, 
$R_{\mathrm{sh}}\propto
t^{\lambda}$ and hence $u_{\mathrm{sh}}\propto t^{\lambda-1}$. The maximum
momentum of accelerated particles in the high velocity limit (7) has the
scaling $p_{\max}\propto u_{\mathrm{sh}}^{2}R_{\mathrm{sh}}\rho^{1/2}\propto
t^{\lambda(3-\frac{s}{2})-2}\propto R_{\mathrm{sh}}^{3-\frac{s}{2}-\frac
{2}{\lambda}}$, so that $R_{\mathrm{m}}(p)\propto 
p^{\frac{1}{3-\frac{s}{2}-\frac{2}{\lambda}}}$.
Now Eq. (25) at $\lambda<4/(6-s)$ gives the
following shape of the average spectrum of run-away particles:
\begin{equation}
Q(p)\propto p^{-4-\frac{\lambda(5-s)-2}{2-\lambda(3-\frac{s}{2})}}. 
\label{27}
\end{equation}
A characteristic of the adiabatic regime is the relation $\lambda=\frac
{2}{5-s}$ and therefore Eq. (27) gives $Q(p)\propto p^{-4}$ in agreement 
with
Eq. (26). The Chevalier - Nadyozhin solution for the ejecta-dominated stage
has $\lambda=\frac{k-3}{k-s}$. With the set of parameters excepted in 
Section
3, we have then $Q(p)\propto p^{-6.5}$ for the acceleration at the shock
produced by the Type II SN in the presupernova star wind ($s=2$, $k=10$,
$\lambda=7/8$), and $Q(p)\propto p^{-7}$ for the acceleration at the shock
produced by the Type Ia SN in the uniform interstellar medium ($s=0$, $k=7$,
$\lambda=4/7$). Thus the cosmic rays accelerated at the ejecta dominated 
stage
have higher energies than at the later adiabatic stage but the average 
energy
spectrum of produced cosmic rays is rather steep at the ''canonical'' choice
of presupernova star parameters.

%figure2

\begin{figure*}[bthp]
\centering
\includegraphics[width=17.0cm]{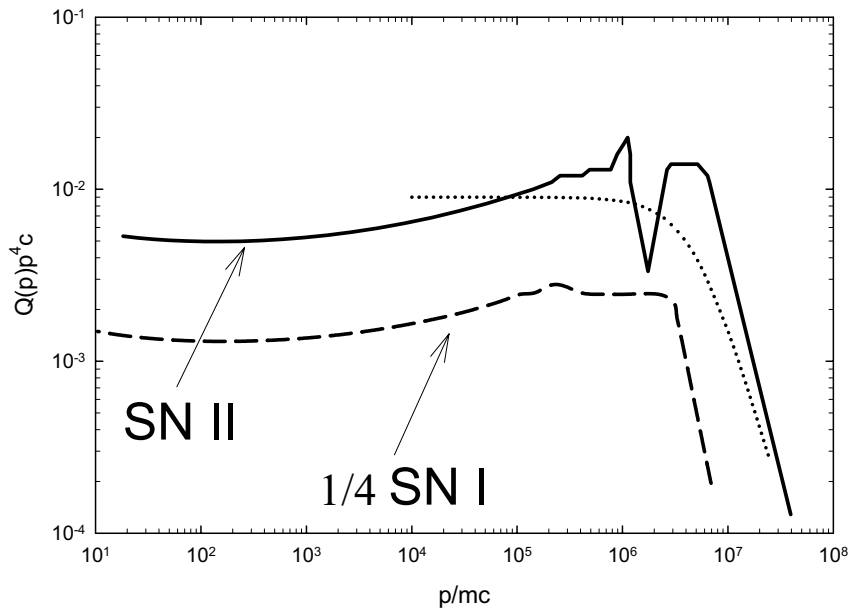}
\caption{The solid line shows the average source spectrum $Q(p)p^{4}c$ 
(given
in units $\frac{\xi_{\mathrm{cr}}}{0.5}\nu_{\mathrm{sn}}\mathcal{E}$ per
steradian)\ for protons released into the interstellar medium during SNR
evolution after SNII explosion in the wind of RSG progenitor star. The 
dashed
line presents the case of SNIa explosion in the uniform interstellar gas; 
the
average source spectrum is multiplied by 1/4. The dotted line shows the 
shape
of proton source spectrum used by H\"{o}randel (2003) to fit the KASCADE 
data.}
\label{Fig2}
\end{figure*}

The results of our numerical calculations of the average spectra for Type II
and Type Ia SNe are shown in Fig. 2. The parameter $\varkappa$ is equal to 
$0.1$
in a high-velocity regime (7) and it is equal to $0.04$ in a low-velocity 
regime (9).

The calculations for Type II SN are based
on Eqs. (5), (6), (15), (17) and (25). For the set of parameters accepted in
the present paper, the Type II supernovae are able to accelerate cosmic ray
protons to the maximum energy of the order $4\times10^{16}$ eV if the
acceleration starts one week after the SN explosion when $u_{\mathrm{sh}
}\approx2.4\times10^{4}$ km/s. The energy spectrum is close to $p^{-4}$ at
energies less than about $6\times10^{15}$ eV and it experiences the 
steepening
above this energy. Thus the proton knee lies
at about $6\times10^{15}$ eV in good agreement with the observational data. 
The
sharp dip in the average proton spectrum at
$p/mc\sim1\times10^{6}-3\times10^{6}$ is caused by the assumed abrupt change
of the gas density at the boundary between the dense Red Super Giant wind 
and the
low density bubble. We run calculations up to the maximum shock radius $60$ 
pc
(the corresponding SNR age is $9\times10^{4}$ yr) when the Mach number 
approaches
$3$ and the use of the particle spectrum (1) characteristic of the strongly
modified shocks can not be longer justified. The protons are accelerated to 
about
20 GeV at this moment.

The calculations for Type Ia SN in Fig. 2 are based on
Eqs. (5), (6), (14), (16) and (25).
The calculations were made for the shock radius range from $0.2$ to $30$ pc
(the SNR age from $4$ yr to $1.3\times10^{5}$ yr). The shock velocity is 
changing
during this period from $2.8\times10^{4}$ km/s to $91$ km/s. The protons are
accelerated from the maximum energy $7\times10^{15}$ eV to about $10$ GeV.
The approximate proton spectrum
$p^{-4}$ extends to the knee at about $3\times10^{15}$ eV.

The average source spectrum
produced by Type Ia SNe is multiplied by $1/4$ in Fig. 2 that corresponds to 
their
relative burst rate and hence reflects the relative
contribution of this type of supernovae to the total production of cosmic 
rays
in the Galactic disk as compared to the contribution of Type II SNe.

\section{Consistency with Cosmic Ray Data and Discussion}

The spectrum of high-energy cosmic rays in the Galaxy is of the form 
$f\propto
p^{-\gamma},\;\gamma=\gamma_{\mathrm{s}}+b$ under the steady state 
conditions
when the action of cosmic ray sources (with the source power $Q\propto
p^{-\gamma_{\mathrm{s}}}$) is balanced by the escape of energetic particles
from the Galaxy (with the escape time $T\propto p^{-b}$). The observed at 
the
Earth all-particle spectrum of cosmic rays is close to $f\propto p^{-4.7}$ 
at
energies $E\gtrsim10$ Gev/nucleon with a characteristic transition (the
knee, Kulikov \& Khristiansen 1958) ranging across less than one decade in 
the
vicinity of $4\times10^{15}$ eV to the another power-law $f\propto 
p^{-5.1}$.
The last extends to about $5\times10^{17}$ eV where the second knee with the
break $\delta\gamma\sim0.3$\ is seen in the cosmic ray spectrum, see
H\"{o}randel (\cite{horandel}) for review. This structure is usually 
associated with a
severe decrease of the efficiency of cosmic ray acceleration or/and
confinement in the Galaxy. The extragalactic component of cosmic rays 
probably
dominates at $E\gtrsim3\times10^{18}$ eV (Gaisser et al. \cite{gaisser}). In 
the
alternative interpretation (Berezinsky et al. \cite{berezinsky04}),
the Galactic component
falls steep (with $\gamma\sim6$) at $E\gtrsim10^{17}$ eV and the 
extragalactic
component dominates from energy $\sim3\times10^{17}$ eV and on.

The exponent $b=0.3...0.7$ was obtained from the data on the abundance of
secondary nuclei at energies $10^{9}$ to $10^{11}$ eV/nucleon. The secondary
nuclei are produced in cosmic rays in a course of nuclear fragmentation of
more heavy primary nuclei moving through the interstellar gas. The 
uncertainty
in the value of $b$ is mainly due to the choice of specific model of cosmic
ray transport in the Galaxy, see Ptuskin (\cite{ptuskin01}). Hence it 
follows that the
source exponent below the first knee lies in the range $\gamma_{\mathrm{s}
}=4.0...4.4$. The value $\gamma_{\mathrm{s}}\approx4.0$ for the average 
source
spectrum was obtained above in the consideration of particle acceleration by
SNR shocks during their adiabatic evolution (though smaller $b\sim0.3$ and
consequently larger $\gamma_{\mathrm{s}}\sim4.4$\ would be more favorable 
for
the explanation of high isotropy of cosmic rays observed at $10^{12}$ to
$10^{14}$ eV). According to the results of Section 4, the calculated average
source spectrum $p^{-4}$\ for protons accelerated by a ''typical'' Type II 
SNe
extends up to about $6\times10^{15 { }}$eV that coincides with the
observed position of the knee $\sim4\times10^{15}$ eV within the accuracy
of our analysis. The knee position at 3-5 PeV was determined in the recent 
KASCADE
experiment (H. Ulrich et al. \cite{ulrich}).
The scaling
of the knee position in our model is $p_{\mathrm{knee}}\propto Z\varkappa
\xi\mathcal{E}\dot{M}^{1/2}M_{\mathrm{ej}}^{-1}u_{\mathrm{w}}^{-1}$
for the explosion in the stellar wind and $p_{\mathrm{knee}}\propto
Z\varkappa\xi\mathcal{E}M_{\mathrm{ej}}^{-2/3}n_{0}^{1/6}$ for the explosion
in the uniform interstellar medium.

As was reminded earlier, the diffusive shock acceleration at the strong
nonmodified shock produces the spectrum $f\propto p^{-4}$. The back reaction
of efficiently accelerating particles modifies the shock structure that 
results in a
more flat particle spectrum (see references at the beginning of Section 2
and Eq. (1) where $a\sim0.5$ if the shock modification is very strong).
However, the numerical simulation of acceleration by SNR shocks under the
standard assumption of Bohm diffusion in the shock precursor (calculated for
the interstellar magnetic field strength) and with efficient confinement of
accelerated particles during all SNR evolution gives the overall source 
spectrum
that is close to $p^{-4}$ (Berezhko et al. \cite{berezhko96}).
Berezhko \& V\"{o}lk (\cite{berezhko00}) pointed out that the last result is 
in
some sense accidental. The late stages of the SNR evolution are important 
here
since relatively weak
shock produces steep particle spectrum that has an effect on the overall 
spectrum.
The situation is different in the model
discussed in the present paper because the coefficient of diffusion is
strongly decreasing with SNR age and the cosmic rays with energies larger 
than
$10-30$ GeV/nucleon leave the supernova shell as the run away particles when
the shock remains strong. The final average source spectrum of high-energy 
cosmic
rays with energies larger than $10-30$ GeV/n is close to
$p^{-4}$ provided that the shock evolution is approximately adiabatic and 
the
efficiency of particle acceleration $\xi_{\mathrm{cr}}$ is roughly constant.
The source spectrum of particles with energies less than $10-30$ Gev/n may 
be more
steep because they are accelerated by not very strong shocks. In this 
connection
it should be noted that the source
spectrum in the basic empirical model of cosmic ray propagation in the 
Galaxy is of the form
$Q(p)p^{2}\propto p^{-2.4}$ at $E<30$ GeV/n, and $Q(p)p^{2}\propto 
p^{-2.15}$ at $E>30$ GeV/n,
see Jones et al. (\cite{jones}).

There are other important differences between the standard and the presented 
here
models of cosmic ray acceleration.
As noted before, our
model of cosmic ray acceleration with strong increase in time of the 
diffusion
coefficient and the corresponding decrease of maximum particle energy may
naturally explain why the SNRs are generally not bright in very high energy
gamma-rays at the age larger than a few thousand years. At this period of
time, there are no particles with energies needed to generate the very high
energy gamma-rays in SNR shell. Another problem is the contribution of
gamma-ray emission from numerous unresolved SNRs with relatively flat 
spectra
to the diffuse galactic background at very high energies. Working in the
standard model Berezhko \& V\"{o}lk (\cite{berezhko03}) took the maximum
possible energy of
protons accelerated in SNRs equal to $10^{14}$ eV, i.e. well below the knee
position, and it allowed not to exceed the upper limits on the diffuse 
gamma-ray
emission at $4\times10^{11}$ to $10^{13}$ eV obtained\ in the Whipple, 
HEGRA,
and TIBET experiments. In the model considered in the present work even with
the efficient proton acceleration that may go beyond the knee, the expected
gamma-ray emission from SNRs at $4\times10^{11}$ eV is order of
magnitude smaller than in the model with Bohm diffusion. For a similar 
reason,
the standard model compared to the present model predicts larger ratio of 
fluxes of
secondary and primary nuclei formed at very high energies through the
reacceleration of secondaries by strong shocks and through the direct
production of secondaries by primary nuclei with flat energy spectra inside
SNRs, see Berezhko et al. (\cite{berezhko03}).

The interpretation of energy spectrum beyond the knee in the present model 
is
associated with the cosmic ray acceleration during the ejecta dominated 
stage
of SNR evolution when the protons gain by an order of magnitude larger
energy than at adiabatic stage but the number of particles involved in the 
shock
accelerated is relatively small. The average source spectrum of accelerated
particles is not universal at this stage. It has a power law high-energy
asymptotics with the exponent $\gamma_{\mathrm{s}}$\ which value is very
sensitive to the parameter $k$. The last is not well determined from the
observations but the typical values accepted in our calculations were $k=10$
for the Type II SN explosion in the wind of a Red Super Giant progenitor, 
and
$k=7$ for the Type Ia SN explosion in the uniform interstellar medium \ (see
Chevalier \& Liang \cite{chevalier89}) that give $\gamma_{\mathrm{s}}=6.5$ 
and
$\gamma_{\mathrm{s}}=7$ respectively, see Section 4. To illustrate
the range of possible uncertainty, it is worth noting that the value $k=5.4$
was suggested for the Type Ia SNe by Imshennik at al. (\cite{imshennik}).
This value of
$k$ results in $\gamma_{\mathrm{s}}=$ $4.3$ at the ejecta dominated stage.

The breaks and cutoffs in the spectra of ions with different charges should
occur at the same magnetic rigidity as for protons, i.e. at the same ratio
$p/Z$ (or $E/Z$ for ultrarelativistic nuclei). The data of KASCADE 
experiment
(Ulrich et al. \cite{ulrich}) for the most abundant groups of nuclei 
(protons, helium,
CNO group, and the iron group nuclei) are, in general, consistent with this
concept. According to H\"{o}randel (\cite{horandel}) the good fit to the 
observations is
reached if an individual constituent ion spectrum has a gradual steepening 
by
$\delta\gamma\sim2$ at energy $4\times10^{15}Z$ eV. Eq (27) shows that the
value $\delta\gamma_{\mathrm{s}}=2$ can be obtained at $k=9$, $s=2$ (the SN
explosion in the progenitor wind), or $k=6.6$, $s=0$ (the SN explosion in 
the
uniform interstellar medium) that is not very different from our accepted
''typical'' values, see Fig. 2.

At present, the main problem of the data interpretation centers around the
second knee in the cosmic ray spectrum. The natural assumption that all
individual ions has only one knee at $\sim4\times10^{15}Z$ eV and that the
knee in the spectrum of iron ($Z=26$) expected at about $10^{17}$ eV
explains the second knee in the all-particle spectrum does not agree with 
the
observed position of the second knee at $5\times10^{17}$ eV. One way out was
suggested by H\"{o}randel (\cite{horandel}) who included all elements up to 
$Z=92$ into
the consideration and assumed that $\gamma_{\mathrm{s}}$ decreases with $Z$ 
to
rise the contribution of ultra heavy nuclei from Galactic sources to the
cosmic ray flux at $\gtrsim10^{17}$ eV. Of considerable promise is the
approach by Sveshnikova (\cite{sveshnikova}) who took into account the 
dispersion of
parameters of SN explosions in her calculations of the knee position and the
maximum particle energy. It leads to the widening of the energy interval
between the two knees in the overall all-particle spectrum. This analysis
should be supplemented by the account of different chemical composition of 
the
progenitor star winds that determines the composition of accelerated cosmic
rays (Silberberg et al. \cite{silberberg}). We plan the future work on this 
topic in the
frameworks of the model developed in the present paper and Paper I. It 
should
be noted that the model by Berezinsky et al. (\cite{berezinsky04})
is quite consistent with
the falling down of the flux from Galactic sources above $10^{17}$ eV
since the conjunction with the intergalactic cosmic ray flux in their model
occurs at relatively low energy.

There is also a very different scenario which assumes the strong 
reacceleration of cosmic rays
above the knee by the collective effect of multiple SNR shocks in the 
violent
regions of Galactic disk (Axford \cite{Axford}, Bykov \& Toptygin 
\cite{Bykov}, Klepach et al. \cite{Klepach})
or Galactic wind (V\"{o}lk \& Zirakashvili \cite{Zir2004}).

Finally, it is worth noting that in principle the knee may
arise not in the sources but in the process of cosmic ray propagation in the 
Galaxy,
e.g. as a result of interplay between the ordinary and the Hall diffusion
(Ptuskin et al. \cite{drift}, Roulet \cite{roulet}).
However, this explanation requires the existence of the power-law source 
spectrum
which extends without essential breaks up to about $10^{18}$ eV or even 
further.

\section{Conclusion}

The accounting for non-linear effects which accompany the cosmic ray 
streaming
instability raises the maximum energy of accelerated particles in young SNRs
above the standard Bohm limit by about two orders of magnitude. It also 
considerably
reduces the maximum energy of particles that are present inside SNRs at the 
late Sedov stage  
if, as it was assumed in our calculations in Section 4, the 
cosmic ray diffusion coefficient downstream of the shock is not much smaller 
than the diffusion coefficient in the cosmic ray 
precursor of the shock and the 
energetic particles with $p\varpropto p_{\max }$ runs away from the 
SNR interior. 
In the present paper we studied the effect of arising strong time
dependence of maximum particle momentum $p_{\max}(t)$ on the average 
spectrum
of cosmic rays injected into interstellar space from many supernova remnants
over their lifetime. The instantaneous cosmic ray spectrum at strongly
modified shock is flat ($f_{0}\varpropto p^{-4+a},$ $a>0$, Eq. (1)) and the
particle energy density is mainly determined by the particles with maximum
momentum $p_{\max}(t)$. The instantaneous source spectrum of the run away
particles is close to the delta function ($q_{\mathrm{ra}}(t,p)\propto
\delta(p_{\max}(t)-p)$, Eq. (24)). At the same time, the assumption that the
constant fraction $\xi_{cr}$ of incoming gas momentum flux goes to the
cosmic ray pressure at the shock, and the fact that the supernova remnant
evolution is adiabatic lead to the average on ensemble of SNRs source 
spectrum
of ultrarelativistic particles that is close to $Q_{\mathrm{ra}}\varpropto
p^{-4}$ from energies $10-30$ GeV/n up to the knee position in the observed 
cosmic ray spectrum
independent of the value of $a$, see Eq. (26) and Fig 2. This source 
spectrum is consistent with the
empirical model of cosmic ray propagation in the Galaxy. The acceleration at
the preceding ejecta-dominated stage of SNR evolution provides the steep 
power-law
tail in the particle distribution at higher energies up to
$\sim10^{18}$ eV (if the iron nuclei dominate at these energies).
The knee in the observed energy spectrum of cosmic rays at 
$\sim4\times10^{15}$ eV is explained in our model by the
transition from the particle acceleration at the ejecta-dominate stage to 
the adiabatic stage of SNR shock evolution.
In spite of approximate character of our consideration, it seems that the 
suggested
scenario of particle acceleration can explain the energy spectrum of 
Galactic
cosmic rays.

\begin{acknowledgements}
The authors are grateful to Evgenij Berezhko and
Heinz V\"{o}lk for fruitful discussions. We thank the referee for
valuable comments. This work was supported by
the RFBR grant at IZMIRAN. VSP was also supported by the NASA grant
during his visit to the University of Maryland where part of this work was
carried out.
\end{acknowledgements}
\appendix
\section{Thin shell approximation}

The thin-shell approximation can be used when the swept-up gas is 
concentrated
in a thin layer behind the shock. In particular, it is applied to the case 
of
a spherical adiabatic shock, see Ostriker J.P. \& McKee C.F. 
(\cite{ostriker}) and
Bisnovatyi-Kogan \& Silich (\cite{kogan}) for detail. The total mass of the 
gas shell
involved in the motion and confined by the shock of radius $R_{\mathrm{sh}}$
in the spherically symmetrical case is
\begin{equation}
M=M_{\mathrm{ej}}+4\pi\int_{0}^{R_{\mathrm{sh}}}drr^{2}\rho(r), \label{A1}
\end{equation}
where $M_{\mathrm{ej}}$ is the ejected mass,$\ \rho$ is the density of 
ambient gas.

The equation of momentum conservation is
\begin{equation}
\frac{d(Mu)}{dt}=4\pi R_{\mathrm{sh}}^{2}\left(  P_{\mathrm{in}}-P\right)  .
\label{A2}
\end{equation}
Here $u$ is the gas velocity behind the shock, $P_{\mathrm{in}}$ is the
pressure behind the shock, and $P$ is the pressure of ambient gas. For the
adiabatic blastwave, $u$ is related to the shock velocity $u_{\mathrm{sh}
}=\frac{dR_{\mathrm{sh}}}{dt}$ by the equation $u_{sh}=\frac{\gamma
_{\mathrm{ad}}+1}{2}u$, where $\gamma_{\mathrm{ad}}$ is the ratio of the
specific heats (adiabatic index). The energy of explosion $\mathcal{E=E}
_{\mathrm{th}}+\frac{1}{2}Mu^{2}$ consists of the internal energy
$\mathcal{E}_{\mathrm{th}}=\frac{4\pi R_{\mathrm{sh}}^{3}}{3\left(
\gamma_{\mathrm{ad}}-1\right)  }P_{\mathrm{in}}$\ and the kinetic energy.

Now for the case of very strong shock when $P_{\mathrm{in}}$ can be omitted 
as
negligible compared to $P$, Eq.(A2) can be presented as:
\begin{equation}
\frac{d(Mu)^{2}}{dR_{\mathrm{sh}}}=\frac{12(\gamma_{\mathrm{ad}}-1)}
{(\gamma_{\mathrm{ad}}+1)R_{\mathrm{sh}}}\left(  \mathcal{E}M-\frac{1}
{2}\left(  Mu\right)  ^{2}\right)  . \label{A3}
\end{equation}
The solution of Eq. (A3) allows finding the shock velocity and the shock age
as functions of the shock radius:
\[
u_{\mathrm{sh}}(R_{\mathrm{sh}}) =\frac{\gamma_{\mathrm{ad}}+1}{2}\left[
\frac{2w\mathcal{E}}{M^{2}(R_{\mathrm{sh}})R_{\mathrm{sh}}^{w}}\int
_{0}^{R_{\mathrm{sh}}}drr^{w-1}M(r)\right]  ^{1/2},\label{A4}
\]
\begin{equation}
t(R_{\mathrm{sh}}) =\int_{0}^{R_{\mathrm{sh}}}\frac{dr}{u_{\mathrm{sh}
}(r)},
\end{equation}
where $w=\frac{6(\gamma_{\mathrm{ad}}-1)}{\gamma_{\mathrm{ad}}+1}$ that
coincides with Eq. (17) in the main text.

\end{document}